# Directional free-surface flows on a microfluidic chip with interplay of electrical and thermal fields

Golak Kunti, Jayabrata Dhar, Anandaroop Bhattachariya & Suman Chakraborty[1]

Department of Mechanical Engineering, Indian Institute of Technology Kharagpur,

Kharagpur, West Bengal - 721302, India

[1]*E-mail address* of corresponding author: suman@mech.iitkgp.ernet.in

Free-surface electrokinetic flows have been attracting increasing attention from the research community over recent times, as attributable to their diverse fields of applications ranging from fluid mixing, particle manipulation to bio-chemical processing on a chip. Traditional electrokinetic processes governing free surface flows are mostly effective in manipulating fluids having characteristically low values of the electrical conductivity (lower than 0.085 S/m). Biological and biochemical processes, on the other hand, typically aim to manipulate fluids having higher electrical conductivities ($>0.1\,\text{S/m}$). To circumvent the problem of traditional electrokinetic processes in manipulating highly conductive fluids mediated by free surface flows, here we demonstrate a novel on-chip methodology for the same by exploiting the interaction between an alternating electric current and a thermal field. The consequent local gradients in electrical properties can be tuned to direct the flow towards a specific direction on the interface, opening up a new realm of on-chip process control without necessitating the creation of a geometric confinement. Similar concept of free-surface flow can be used to trigger the fluid along a open-channel flow

**1. Introduction**

Free-surface problems are the studies where location of the field boundary is not known initially. However, coupled solution of the non-linear field equations enables the determination of the field boundary at any location in the domain (Qian *et al.* 2009). In many engineering and geophysical applications, such as flow in open channels, bubbly flows, hydrodynamics, rivers, lakes, boiling, cavitation, sea waves etc. free-surface flow arises profoundly. Recently, free-surface hydrodynamics in micro electro-mechanical systems (MEMS) has potential applications for fluid mixing, chip cooling cleaning, etc.(Dey & Joo 2015). In the literature, investigations of electrohydrodynamically modulated free-surface flow were reported and established a new kind of flow perturbation of applied electric field on liquid with a free surface (Joo 2008; Qian *et al.* 2009).

Over the last decade, technological advances in fabrication methodologies have triggered the emergence of a plethora of new applications for on-chip control of engineering and biological processes (Gao *et al.* 2011; Temiz *et al.* 2015). In many of these applications, alternating current electrokinetics (ACEK) has been successfully deployed, considering its excellent on-chip integrity, low excitation voltage and precise controllability over miniaturized scales(Dalton & Kaler 2007). Efficient and effective fluid mixing (Feng *et al.* 2007; Kunti *et al.* 2017b; Ng *et al.* 2009), pumping(Du & Manoochehri 2010; Kunti *et al.* 2017a; Lian & Wu 2009), particle concentrating (Kumar *et al.* 2011; Kwon *et al.* 2012; Wang *et al.* 2014), multiphase flow (Kunti *et al.* 2017c; 2017d; 2018a; 2018b), and sorting(Martinez-Duarte 2012) have been demonstrated with success, by employing such mechanisms of flow manipulation.

The pertinent operating parameters that control the efficacy of an ACEK process primarily include the frequency of the electrical pulsation and electrical conductivity of the working fluid. Beyond a threshold frequency (>100 kHz (Du & Manoochehri 2008)) and electrical conductivity ($>0.085\,\text{S/m}$(Studer *et al.* 2004)), ACEK is inefficient for fluid manipulation. However, a wider operating regime of the device may be potentially explored, by creating a thermal gradient in the system, via triggering the gradients of the relevant electrical properties with variations in temperature. Due to several intrinsic advantages of manipulating high conductivity fluid, such processes, commonly known as electrothermal (ET) actuation, have been widely used for transportation of fluids in confined geometrical pathways (Du & Manoochehri 2010; Lang *et al.* 2015; Wu *et al.* 2007). The performance of such pumping mechanisms have been successfully demonstrated in widely varying scenarios such as grooved channel (Du & Manoochehri 2008), thin film heater, DC biasing (Lian & Wu 2009), two-phase electrical signalling (Zhang *et al.* 2011) etc.

Demands of deploying geometric confinements such as microchannels may often impose fabrication and implementation restrictions for applying electrothermal flows in conjunction with simple on-chip configurations for practical applications ranging from chemical and thermal processing to biomedical technology. Circumventing these limits, here we report the first results of generating controlled directional free-surface flows on a microfluidic chip by deploying an intricate interaction between the electrical and the thermal field, without necessitating the fabrication of a confined geometric passage. Interestingly, the thermal field exploited for this purpose is not externally imposed but is intrinsically induced through Joule heating. This, in turn, imposes inhomogeneity in conductivity and permittivity, inducing free charges into bulk fluid. These mobile charges effectively set the free-surface fluid into motion. Further, suitable arrangement of the electrodes on the chip may impose designed variations in the temperature gradient over the domain, leading to the inception of free-surface flow in a preferential direction.

## 2. Materials and methods

### 2.1. ACET theory

Electrothermal mechanism is the consequence of AC electric field and induced thermal field. We have considered quasi-electrostatic field (Morgan & Green 2003; Ramos 2011), where magnetic field can be neglected. The distribution of the voltage ($\varphi$) in the frequency domain can be obtained solving the following equation (Castellanos A 1998; Hong *et al.* 2012):

$$\nabla \cdot \left( \sigma \nabla \varphi_r \right) = 0, \qquad (2.1)$$

where $\sigma$ is the electrical conductivity. The subscript r stand for real component.

Electrothermal forces arise from the gradients of electrical properties which are generated by the temperature gradient in the electrolyte. Application of electric field causes Joule heating as:

$$\sigma \langle E^2 \rangle = \rho C_p \mathbf{V} \cdot \nabla T - k \nabla^2 T, \qquad (2.2)$$

where $E\,(=|\mathbf{E}|, \mathbf{E} = -\nabla \varphi)$ is the electric field. $\mathbf{V}$ is the fluid velocity. $\rho$ is the mass density. $C_p$ is the specific heat. $k$ is the thermal conductivity. Generated gradients in conductivity and permittivity are $\nabla \sigma = (\partial \sigma / \partial T) \nabla T$ and $\nabla \varepsilon = (\partial \varepsilon / \partial T) \nabla T$, respectively. For aqueous solutions, these gradients, typically are $(\partial \varepsilon / \partial T)/\varepsilon = \alpha \approx -0.4\% \, \text{K}^{-1}$ and $(\partial \sigma / \partial T)/\sigma = \beta \approx 2\% \, \text{K}^{-1}$ (Lide 2003). Inhomogeneities in electrical properties generate mobile charges in the fluid, as governed by: $q = (\sigma \cdot (\partial \varepsilon / \partial T) - \varepsilon \cdot (\partial \sigma / \partial T))/(\sigma + i\omega\varepsilon) \cdot \nabla T \cdot \mathbf{E}; \, \partial / \partial t = i\omega$. $\omega$ is the frequency of the AC signal (Ramos *et al.* 1998).

Induced mobile charges and electric field combinedly generate electrothermal forces, with the volumetric force density given by (Ramos *et al.* 1998):

$$\mathbf{F}_E = q\mathbf{E} - 0.5\mathbf{E}^2 \nabla \varepsilon. \quad (2.3)$$

The equation governing fluid motion is given by:

$$\rho \partial \mathbf{V} / \partial t = -\nabla p + \mu \nabla^2 \mathbf{V} + \langle \mathbf{F}_E \rangle + \mathbf{F}_b, \quad (2.4)$$

where $\mu$ is the fluid viscosity. Buoyancy force reads $\mathbf{F}_b = \Delta \rho \mathbf{g} = (\partial \rho / \partial T) \Delta T \mathbf{g} = -\rho \beta \Delta T \mathbf{g}$; $\Delta T = T - T_a$. $T_a$ is the ambient temperature and $\beta = -0.0002 \text{K}^{-1}$ (Du & Manoochehri 2008) is the thermal expansion coefficient.

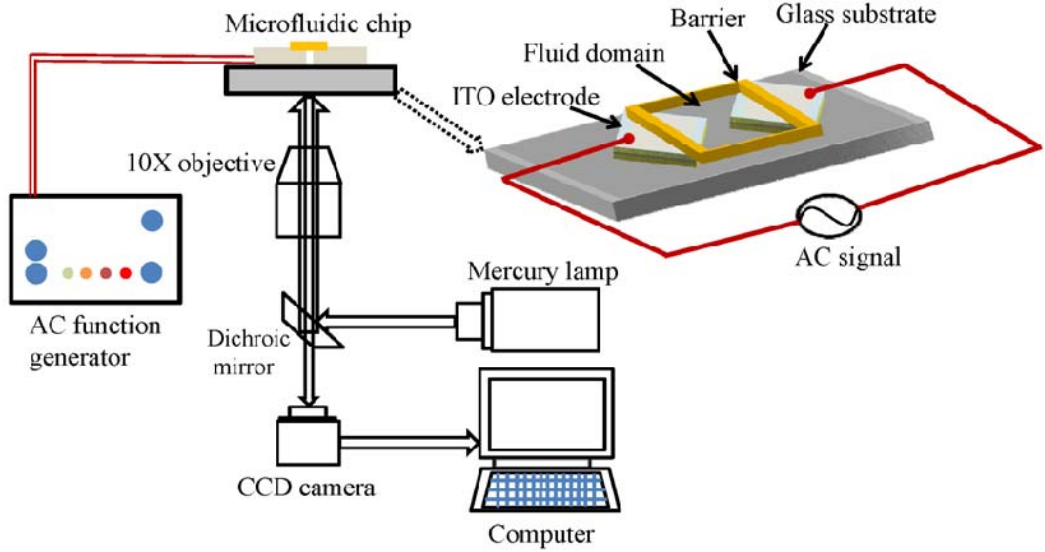

FIGURE 1. Schematic diagram of experimental set up of free-surface flow. Fluid domain is filled by KCl solution. On activation of electric filed a hot spot is generated at the centre of the fluid domain and fluid flows towards the hot spot.

## 2.2. Experimental method

The concept geometry is shown in Figure 1, where, for simplicity, we have taken two pieces of indium-tin-oxide (ITO) electrodes. While deployment of more numbers of electrodes with tunable spacings may impose more intricate control over the free surface flow, we do not attempt to elucidate those additional results here for the sake of brevity. In the demonstrative example we first refer to, the minimum gap between the electrodes is 65μm and is increasing from the centre to the edges of the electrodes. The electric field is maximum at the narrow gap and it gradually decreases towards the edges of the electrodes as the electric field strength decreases with increasing electrode spacing. As a result, Joule heating caused by the electric field generates temperature gradient which decreases from narrow gap to wider gap. Accordingly, fluid moves towards the narrow gap of the electrodes. Glasses with ITO electrodes (thickness: 1.1 mm) were kept on a clean glass substrate. To maintain the gap between the electrodes and to make a barrier, we used a sealing element around the electrodes (see Fig 1). The barrier is far away from the narrow gap so that end effects do not disturb the flow field. KCl (Merck Life Science Pvt. Ltd.) electrolyte solution of electrical conductivity 0.099-1.92 S/m was used as fluid medium. AC field was employed using a function generator (33250A, Agilent). Fluorescence microscopy imaging technique was adopted to track the fluorescent particles (diameter 1 μm, FluoSpheres@carboxylate,

Life Technologies Pvt. Ltd) motion. An inverted microscope (IX71, Olympus) and A CCD camera (ProgRes MFcool) were used to observe and to record the fluid motion, respectively.

## 2.3. Numerical method

The physical system and boundary conditions for numerical simulations is shown in the Figure 2(*a*). Three domains: air, water and glass are coupled for the solution of electric, thermal, and velocity fields. Thickness of the ITO electrode is negligible compared to the other dimension and is not considered in the simulation. Electric and thermal field involve the whole domain where velocity field involves only liquid (i.e., KCl solution) due to low density of air. The width, depth, and height of the simulation domain are 1.479 mm, 1.414 mm, and 3.1 mm, respectively. We confirmed that the size of the domain is sufficiently large to avoid boundary effects. The height of the air, water, and glass are 1 mm, 1.1 mm, and 1 mm, respectively. Three-dimensional simulations were performed to obtain the electric, thermal and velocity fields using Finite Element Method (FEM) based software COMSOL. We discretize the simulation domain using 584554 elements, where meshes are densely packed near the tip of the electrodes. To avoid the field singularity at the corners of the electrodes we consider corner radius of $10\,\mu m$ at these edges. We have applied an AC potential ($\pm\varphi_{rms}$) on the electrodes at a frequency of $\omega$. Outer boundaries are electrically insulated. At the interfaces of fluid-solid and fluid-fluid current continuity is applied. The dielectric constant of water, air, and glass are 80, 1, and 3.1, respectively. Due to low conductivities of air and glass Ohmic currents are neglected in these layers. To obtain the temperature field, one can apply Newton's law of cooling (Dey & Joo 2015) $-k\nabla T \cdot \mathbf{n} = h(T - T_a)$ on the free surface. Here, $\mathbf{n}$ is unit outward normal vector and $h$ is the heat transfer coefficient of air. However, a negligible changes was found when conduction in the air was applied. Thus, continuity of heat flux is applied on the interfaces, instead of applying Newton's law of cooling. On the outer boundaries ambient temperature is set. Free surface boundary condition is imposed on the liquid-air interface. In Figure 2(*a*), $\boldsymbol{\tau}$ denotes the hydrodynamic stress tensor. $\mathbf{t}$ is the unit tangential vector at water surface.

Free-surface flow driven by the electrothermal effect, involves a number of multi-physics phenomena which occur simultaneously. In order to impose the electric field over the simulation domain first, we have solved Eq. (2.1). Eqs. (2.2) and (2.4) are solved simultaneously since these are coupled each other. The resultant force magnitude is used to obtain the directed free-surface flow.

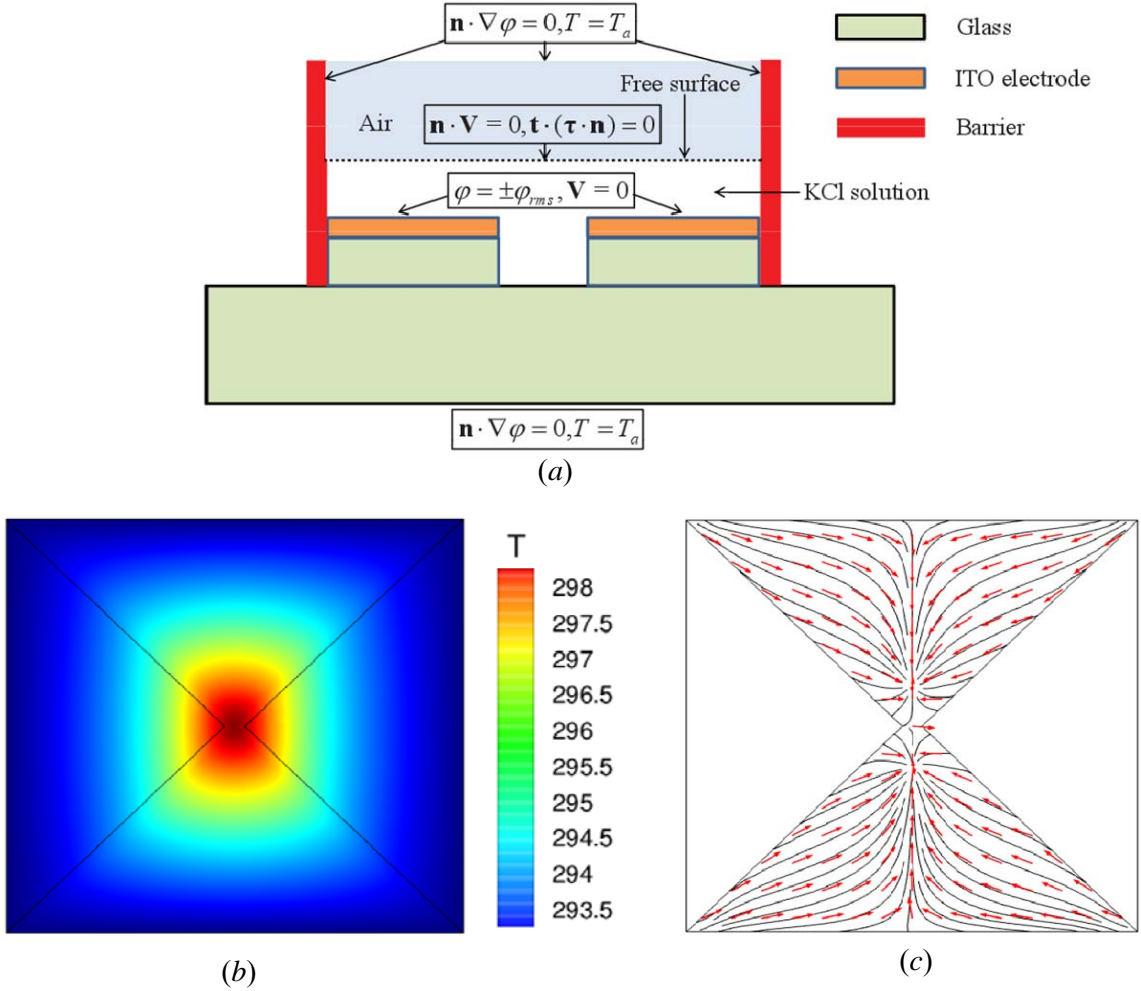

FIGURE 2. (*a*) Depicts simulation domain and boundary conditions. Sectional top view (50 μm below the electrode surface surface) of (*b*) surface plot of temperature and (*c*) streamline contours for electrical conductivity 0.36 S/m, peak voltage 10 V and frequency 1 MHz.

## 3. Results and discussion

In this section, we have discussed influence of operating parameters on controlled directional free-surface flow. Figures 2(*b*) and 2(*c*) show the sectional view (50 μm below the electrode surface) of surface plot of temperature and streamline contours for electrical conductivity 0.36 S/m, peak voltage 10 V and frequency 1 MHz. Electric field is maximum at the narrow gap of the electrodes. Further, one can see that temperature is highest at the narrow gap of the electrodes. The present configuration deals with low Reynolds number flow where inertial effects are less compared to the other forces. Hence, in the energy balance, thermal conduction scales with the Joule heating source, so that $\Delta T \sim E^2$. Therefore, high local temperature gradients prevail at locations having high local electric field strength. It may also be noted that the temperature sharply drops from the narrow gap to the wider gap. This implicitly induces high temperature gradients within the system. This strong thermal field is effective to generate strong electrothermal forces. Fluid motion take places from the cooler region to hotter region. Corresponding streamline plot with arrow shows the flow field for the directional free-surface flow. It is clear that fluid flows towards the narrow gap between two electrodes. Arrows depict the flow direction. In the flow domain, vortices are not found. Therefore, net effective fluid flow can be achieved towards the centre of the electrode system. An important to be mentioned at this juncture that this directional flow towards the centre of the electrode system takes place below the free-surface. To maintain the mass conservation another flow from centre to outer direction is observed at the free surface.

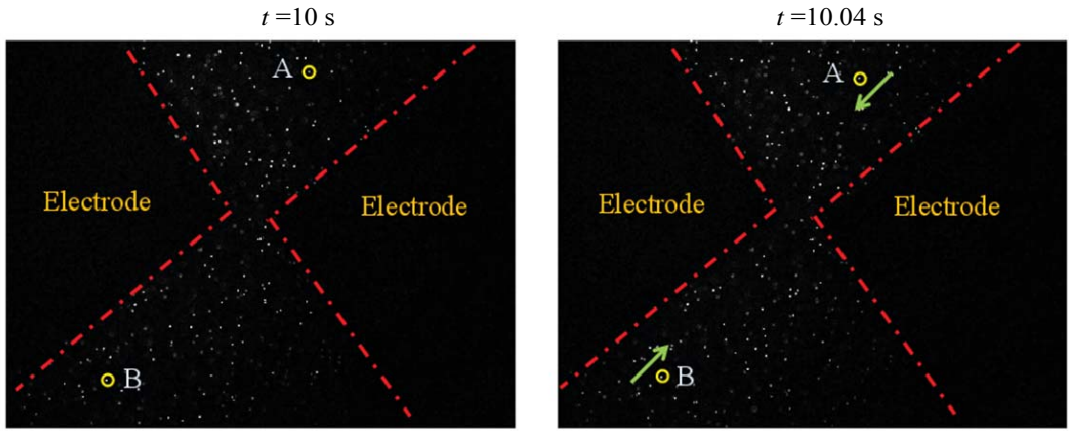

FIGURE 3. Images show tracking of particles to measure fluid velocity and to show the direction of the flow. The operating parameters are electrical conductivity: 0.36 S/m, peak voltage: 10 V, and frequency: 1 MHz.

To show the flow direction, we also represent experimental results of particle tracking in Figure 3. Directions of the particles A and B are shown by arrow. Both particles move towards the centre of the flow domain where electric field and thermal field are strongest. (supplementary movie1 shows the fluid motion for this scenario). A good agreement between the experimental results and numerical predictions may be observed.

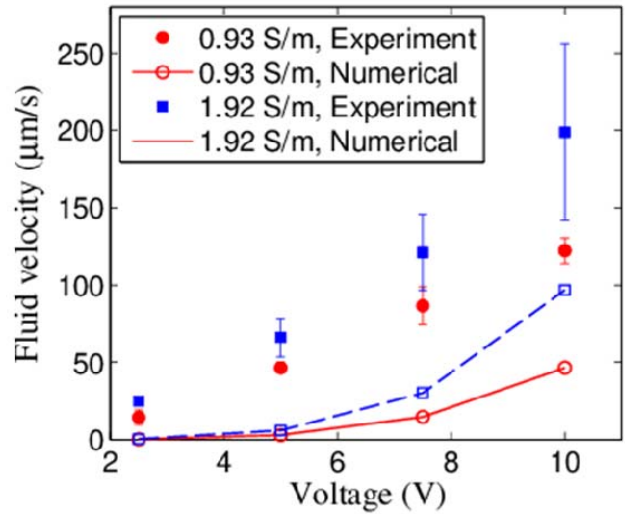

FIGURE 4. Variation in averaged fluid velocity with applied voltage for two different conductivities $\sigma = 0.93, 1.92$ S/m at frequency 1 MHz.

Our experiments have demonstrated that the flow velocity is strongly dependent on applied voltage and solution conductivity. Figure 4 highlights the variation in fluid velocity with the applied voltage for conductivity values of 0.93, 1.92 S/m, whereas the frequency of electrical pulsation is kept at 1 MHz. Here, we have taken average flow velocity of the tracer particles. It is clear that the flow velocity drastically increases with the voltage. Stronger the electric field, higher will be the electrothermal forces. Ideally, since the induced charge is proportional to electric field, the flow velocity $u \propto E^2$. In addition, the thermal field varies with square of the electric field and hence, considering, $u \sim |\nabla T| \times E^2$, we get, from ideal theoretical perspective: $u \propto E^4$. However, from regression analysis of the experimental data of velocity versus voltage for four different conductivity values ($\sigma = 0.099, 0.36, 0.93$ and $1.92$ S/m), the experimentally obtained index of electric field dependence of flow velocity is 2.4, resulting in some deviations from the ideal behaviour. Although the agreement between more realistic numerical solution and experimental data is much closer, there is some deviation from a quantitative perspective. This deviation can be attributed to the fact that some other electrohydrodynamic forces, such as dielectrophoretic effects and induced-charge electroosmosis (ICEO), may alter the flow field albeit to a lesser extent, are not considered in the numerical modeling. On the other hand, during experiment evaporation of the electrolyte may take place, which increases the concentration of the solution. Therefore, at higher conductivity owing to increased concentration leads to higher temperature gradient and higher fluid velocity. In the numerical simulation, conductivity is taken as constant.

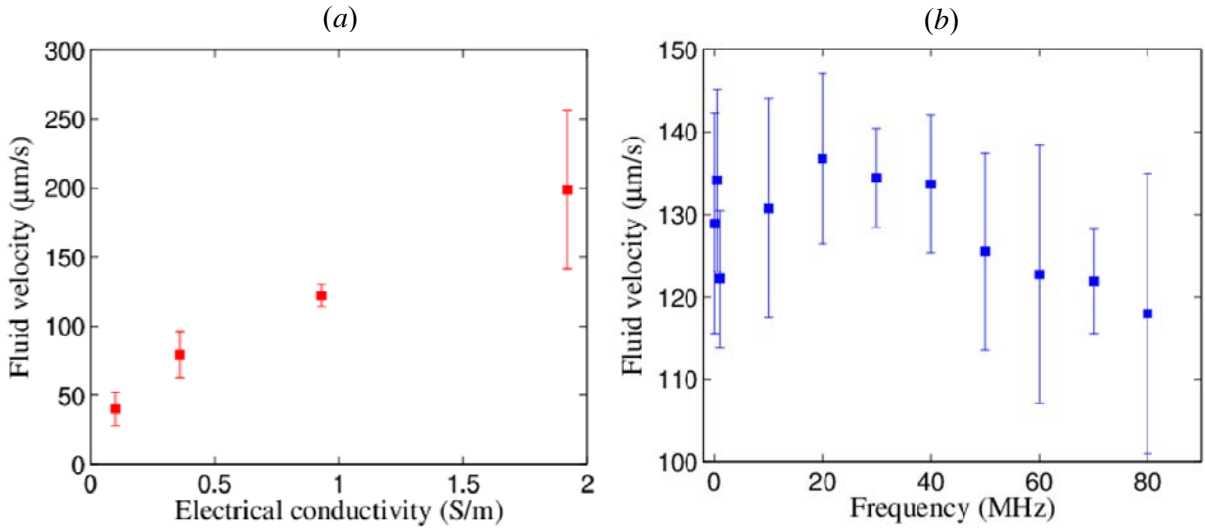

FIGURE 5. (*a*) Fluid velocity as a function of conductivity for peak voltage 10 V and frequency 1 MHz. (*b*) Fluid velocity as a function of frequency for peak voltage of 10 V and conductivity 0.93 S/m.

In Figure 5(*a*), fluid velocity variation with the electrical conductivity of the solution, for applied voltage 10 V and frequency 1 MHz, is presented. It is noted that fluid velocity varies almost linearly with the electrical conductivity of the solution.

One can see that in the energy equation, the temperature gradient varies as $|\nabla T| \propto \sigma$. Since from force balance it was found that fluid velocity is proportional with $\nabla T$, it is expected that fluid velocity has linear dependence with electrical conductivity. Figure 5(*b*) depicts the influence of the frequency on the fluid velocity, for conductivity of 0.93 S/m and voltage of 10 V. The pertinent electrokinetic forces comprise Coulomb force and dielectric force. Coulomb force directly depends on operating frequency whereas dielectric force is independent of the frequency. These two forces oppose each other and form an angle between the relevant force vectors in the phasor space. A crossover frequency, at which the magnitudes of the two forces are equal, may be obtained as: $f_c = \sigma / 2\pi\varepsilon \sqrt{(1-2\beta/\alpha)}$. Thus, it depends on the physical property of the fluid medium. For conductivity $\sigma = 0.93$ S/m, the crossover frequency becomes 718.5 MHz. Below this frequency, the Coulomb force dominates over the dielectric force and owing to complete saturation of free charges the net force almost becomes invariant with the frequency. Our experiments were conducted in the frequency range of 0.1 to 80 MHz. Frequencies beyond 80 MHz were not imposed due to limitation of availability of operating frequency of the function generator. The adopted highest frequency is quite below that of the crossover frequency. Accordingly, the net force in the free surface flow remains almost constant. The measured velocity data shows imperceptible variations in the mean flow velocity, with the lowest and highest velocities in the tune of 118 μm/s and 136 μm/s, respectively. Therefore, alteration of fluid velocity with AC frequency agrees well with the theoretical predictions.

In the previous paragraphs, we have analyzed the characteristics of directional free-surface flow and effects of various important parameters. It was seen that tuning the directionality of the imposed thermal field it is possible to direct the free surface along a specified location. In the following paragraphs we have shown the application of the concept to direct the fluid through a open channel without necessitating confinement. To form an open channel flow configuration we consider a tapper electrode arrangement whose entry (at wider gap of electrodes) and exit (at narrow gap of electrodes) appear as inlet and outlet of the open channel, respectively.

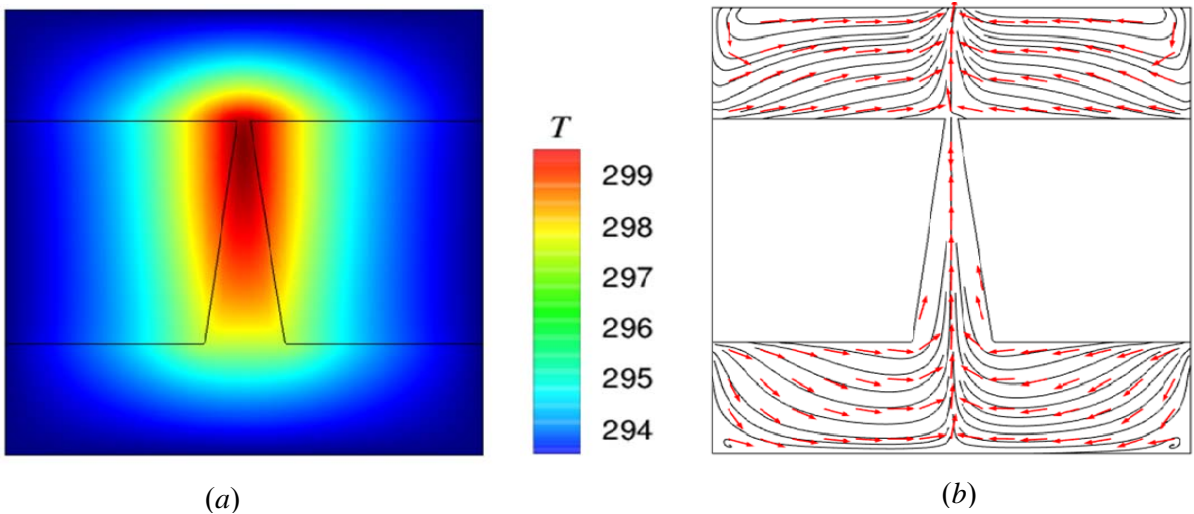

(*a*)          (*b*)

FIGURE 6. Sectional view (50 μm below the electrode surface) of (*a*) temperature (surface plot) and (*b*) streamline contour for peak voltage of 10 V, frequency 1 MHz, and conductivity

0.36 S/m. The spacing's between the electrodes at inlet and outlet are $355\,\mu m$ and $65\,\mu m$, respectively.

The temperature distribution and streamline contours for open-channel flow at depth of $50\,\mu m$ from electrode surface are shown in Figures 6(*a*) and 6(*b*), respectively for applied peak voltage 10 V, conductivity 0.36 S/m and frequency 1 MHz. It is clear that intrinsically induced temperature is highest at the narrow gap and it sharply decreases towards the wider gap. From the energy balance between convection and Joule heating terms, it can be obtained $\rho C_p u_r (\Delta T / l_r) \sim \sigma \varphi_{rms}^2 / l_r^2$. For $\rho \sim 10^3$ kg/m$^3$, $C_p \sim 10^3$ J/kgK, $\Delta T \sim 10$ K, $l_r \sim 10^{-5}$ m, $\sigma \sim 0.1$ S/m, $\varphi_{rms} \sim 1$ V, the order of velocity is $u_r \sim 10^{-3}$ m/s. Therefore, to become dominant of the convection term with other terms in the energy equation the velocity will be order of 1 mm/s. However, our investigations reveal that fluid velocity is lower than threshold value of 1mm/s and hence, conduction term is comparable with Joule heating term. From the balance between conduction dissipation with generated Joule heat results $\Delta T \sim \sigma E^2$. Therefore, temperature is high where electric field strength is high. Narrow gap between the electrodes generates high electric field thereby high temperature. As the spacing between the electrodes increases towards the wider region electric field strength and temperature decreases. This feature causes inhomogeneities in electrical conductivity and permittivity, which, in turn, generate fluid motion via inducing free charges from wider gap to narrow the gap. Streamline contours clearly depict flow field where vortex motion is not seen. Red arrow shows the flow direction which is wider gap to narrow gap. Our experimental results (Figure 7) also show same characteristics of the flow field. We have tracked three particles: (shown by green arrow): particles A, B, and C to show the flow direction (See supplementary movie2 for fluid motion of open-channel flow for electrical conductivity 0.36 S/m, peak voltage 10 V and frequency 10 MHz).

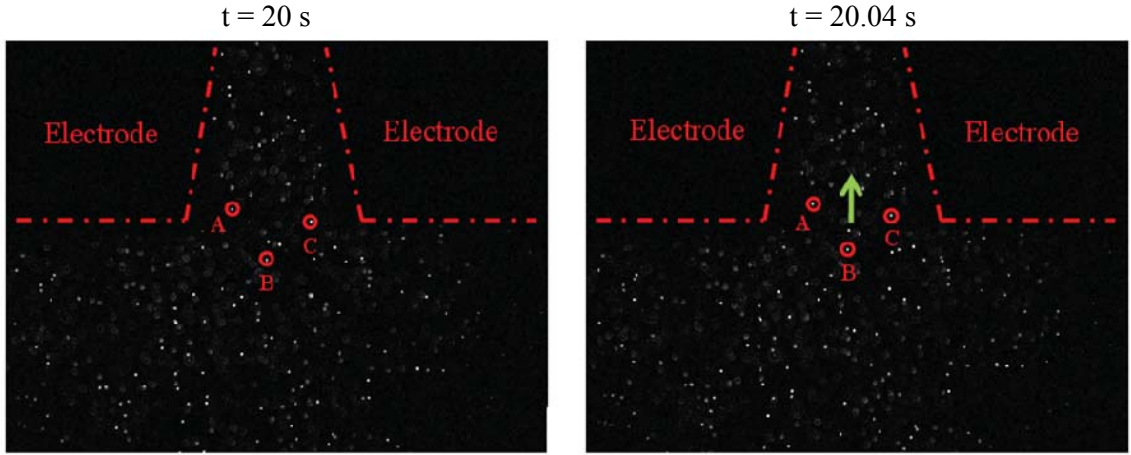

FIGURE 7. Time sequences of images to show the direction of the flow.

To make an assessment of dependency of fluid velocity with electrode spacing and characteristics of tapering we have shown mean fluid velocity at inlet of the open channel for two different cases: (a) widths of the narrow gap $g_o = 65\,\mu m$ and wider gap $g_i = 355\,\mu m$ (b) widths of the narrow gap $g_o = 80\,\mu m$ and wider gap $g_i = 555\,\mu m$ (Figure 8). Subscript i and o stand for inlet and outlet, respectively. For cases (a) and (b) the gap ratios are $g_{r,a} = 5.46$ and $g_{r,b} = 6.93$, respectively. The results are shown as a function of peak voltage (Figure 8(*a*)) and fluid conductivity (Figure 8(*b*)). Although increase in electrode gap from narrow to wider region is sharp for case (b) (since $g_{r,b} > g_{r,a}$) mean fluid velocity is higher for case (a). Fluid velocity depends on how fast electrode spacing becomes narrow to wide. Also, fluid velocity depends on electrode spacing. Both effects are important. In the present scenarios, effect of electrode spacing is much significant than the drastic change in electrode spacing. For case (a) electrode spacings are smaller compared to the case (b). In this case, the electric field strength is higher for case (a) which causes a much sharper temperature gradient to induce large electric forces. As a result, fluid velocity is high in case (a).

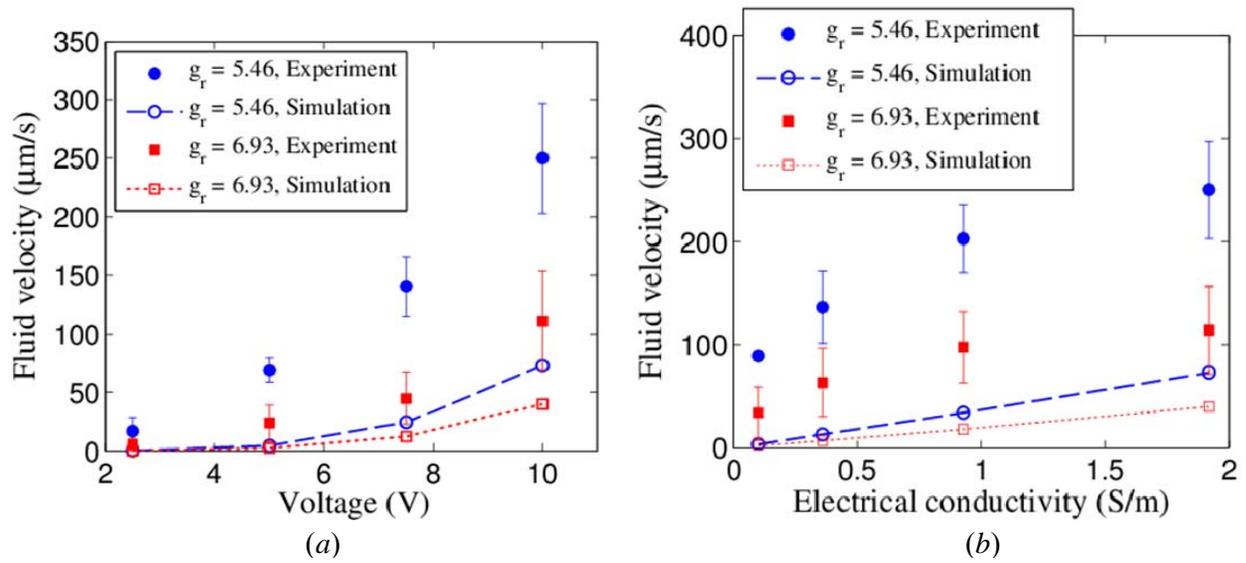

(*a*)          (*b*)

FIGURE 8. Fluid velocity as a function of (*a*) voltage and (*b*) conductivity for two different electrode arrangement: case (a) electrode gap at inlet and outlet $g_i = 355\,\mu m$ and $g_o = 65\,\mu m$; case (b) electrode gap at inlet and outlet $g_i = 555\,\mu m$ and $80\,\mu m$, respectively. For (*a*) other parameters are: conductivity 1.92 S/m and frequency 1 MHz. For (*b*) other parameters are: peak voltage 10 V and frequency 1 MHz.

From Figures 8(*a*) and 8(*b*) it is evident that fluid velocity also functions of applied voltage and solution concentration. Mean fluid velocity drastically increases with voltage and gradually increases with electrical conductivity. At higher voltage strength of the electric field becomes stronger to generate high ACET forces. On the other hand, increased concentration also increases temperature gradient in the fluid domain. Therefore, fluid velocity is increased with ACET forces which is increased with increasing voltage and conductivity. Our numerical data is slightly low compared to measured data. The reason was already stated earlier.

The present understanding of ACET mechanism can be further explored showing the features of the Figure 9 where mean fluid velocity for

$g_r = 1.05$ ($g_i = 220\,\mu m$, $g_o = 210\,\mu m$) and $g_r = 5.46$ ($g_i = 355\,\mu m$, $g_o = 65\,\mu m$) are shown for conductivity 1.92 S/m and frequency 1 MHz. For $g_r = 1.05$, the spacing of the electrodes along the channel is almost same and the temperature gradient is almost uniform. Uniform temperature cannot generate variation in properties and thereby low ACET forces. It is notably that fluid velocity much lower for $g_r = 1.05$ in comparison with the case of $g_r = 5.46$ at which electrode spacing drastically changes from narrow gap to wider gap. Although electrode gap $g_i = 355\,\mu m$ is large compared to electrode gap $g_i = 220\,\mu m$ where electric field strength is high for $g_i = 220\,\mu m$, sharper temperature prevailed in the fluid domain for $g_i = 355\,\mu m$ which results higher fluid velocity. Therefore, tapering of the open channel plays a critical role to alter the flow velocity.

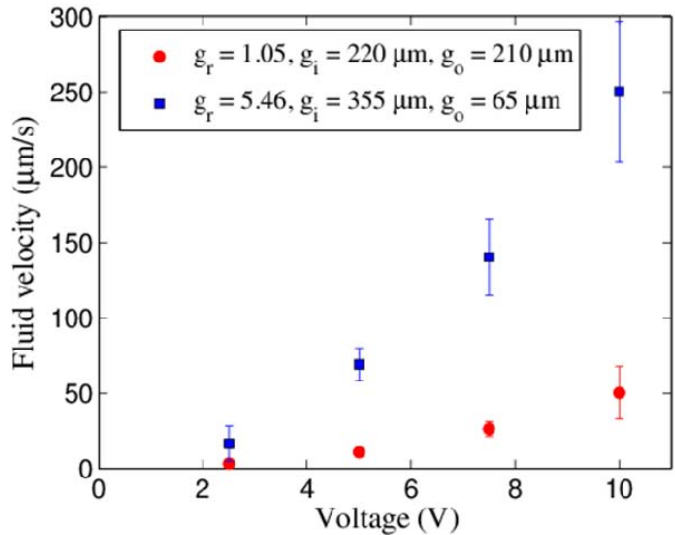

FIGURE 9. Comparison of fluid velocity for two different electrode arrangement: $g_r = 1.05, 5.46$. Parameters used in the experiment are: conductivity 1.92 S/m and frequency 1 MHz.

## 4. Conclusions

In summary, we demonstrate a novel strategy of generating directional free surface flow on a chip, by deploying interactions between electrical forcing and strong local thermal gradients. Using the internally evoked Joule heating, spatially varying gradient in electrical conductivity and permittivity can be generated in the domain. Interactions between gradients in electrical properties and the alternating voltage give rise to mobile charges on the free surface of fluid, setting the fluid also into motion in a specific direction via viscous interactions. Our findings are likely to open up vistas for novel on-chip platforms for biochemical analysis without employing a geometric confinement as well as fundamental theoretical investigations on electro-thermally driven free surface flows.

**Supplementary movies**

Supplementary movies are available at ..................................

This research has been supported by Indian Institute of Technology Kharagpur, India [Sanction Letter no.: IIT/SRIC/ATDC/CEM/2013-14/118, dated 19.12.2013].